\documentclass[runningheads]{llncs}

\def\myauthors{David Otero, Javier Parapar and Álvaro Barreiro}
\def\mytitle{Towards Reliable Testing for Multiple Information Retrieval System Comparisons}

\usepackage{multirow}
\usepackage{soul}
\usepackage{subcaption}
\usepackage{amsmath}
\usepackage{graphicx}
\usepackage{url}
\usepackage[inline]{enumitem}
\usepackage{acronym}
\usepackage{booktabs}
\usepackage{longtable}
\usepackage{pdflscape}
\usepackage{tabularx}
\usepackage{afterpage}
\usepackage{siunitx}
\usepackage{algpseudocode, algorithm}
\usepackage{calc}
\usepackage[hidelinks]{hyperref}
\usepackage{todonotes}
\usepackage{datenumber}
\usepackage[space,sort]{cite}
\usepackage[multiple]{footmisc}
\usepackage[T1]{fontenc}
\usepackage{cleveref}

\author{David Otero\orcidID{0000-0003-1139-0449} \and Javier Parapar\orcidID{0000-0002-5997-8252} \and \'Alvaro Barreiro\orcidID{0000-0002-6698-2946}}

\institute{IRLab, CITIC, Universidade da Coruña, Spain\\ \email{david.otero.freijeiro@udc.es} \email{javier.parapar@udc.es} \email{barreiro@udc.es}}

\authorrunning{Otero et al.}

\hypersetup{
  pdfauthor={\myauthors},
  pdftitle={\mytitle},
  pdfsubject={Information Retrieval},
}

\graphicspath{{plots/}}

\newcommand{\myparagraph}[1]{\paragraph*{\normalsize\bf#1}}

\algblock{Input}{EndInput} 
\algnotext{EndInput}  
\algblock{Output}{EndOutput} 
\algnotext{EndOutput} 
\newcommand{\Desc}[2]{\State \makebox[2em][l]{#1}#2}

\begin{document}

\title{\mytitle}
\titlerunning{Towards Reliable Testing for Multiple IR System Comparisons}

\maketitle

\begin{abstract}

Null Hypothesis Significance Testing is the \textit{de facto} tool for assessing effectiveness differences between Information Retrieval systems. Researchers use statistical tests to check whether those differences will generalise to online settings or are just due to the samples observed in the laboratory. Much work has been devoted to studying which test is the most reliable when comparing a pair of systems, but most of the IR real-world experiments involve more than two. In the multiple comparisons scenario, testing several systems simultaneously may inflate the errors committed by the tests. In this paper, we use a new approach to assess the reliability of multiple comparison procedures using simulated and real TREC data. Experiments show that Wilcoxon plus the Benjamini-Hochberg correction yields Type I error rates according to the significance level for typical sample sizes while being the best test in terms of statistical power.

\keywords{Information Retrieval Evaluation \and Null Hypothesis Significance Testing \and Multiple Comparisons \and Family-Wise Error Rate \and False Discovery Rate}

\end{abstract}


\section{Introduction}
\label{sec:intro}

A thorough evaluation methodology is important to advance in the development of effective retrieval systems. In Information Retrieval (IR), researchers use TREC-like collections to evaluate systems and make inferences about their effectiveness differences. These collections provide a limited number of topics (usually, 50). Null Hypothesis Significance Testing (NHST) is a tool to improve the certainty of those inferences over the few available topics. Statistical tests serve to assess if differences observed in laboratory settings would generalise and hold in operational, real-world settings. Several studies have explored which statistical test is the most reliable for IR evaluation, but most of them are centred around the comparison of just two systems~\cite{Parapar2020,Parapar2021,Urbano2019,Urbano2021,Ferro2022,Smucker2007,Sanderson2005,Webber2008}. Few works have validated multiple testing procedures to compare more than two systems~\cite{Boytsov2013,Ihemelandu2024}.

A typical IR search experiment involves comparing the effectiveness of various systems against several baselines. In this scenario, we are interested in distinguishing the best ones, and thus we need to test every system against all others. If we have $m$ retrieval systems, comparing each one against all others requires $k = \frac{m (m - 1)}{2}$ different pairwise evaluations. With NHST, this means we have a family of $k$ different null hypotheses, one for each possible pair of systems, that we are testing simultaneously\footnote{We may have only one \emph{global} null hypothesis: whether all systems are equal. However, we believe pairwise hypotheses are more interesting in IR evaluation.}. When conducting multiple testing, we need to be very careful. If we make a decision about whether to reject each null hypothesis without accounting for the fact that we have performed several tests, we may end up rejecting many \emph{true} null hypotheses, i.e. having a very high Type I error rate. In terms of new research discoveries, this means that a researcher performing multiple tests without any adjustment that accounts for multiple tests will likely find positive improvements where there may be none in reality. For example, Sanderson and Ferro recently showed that 40\%-50\% of uncorrected significance tests in IR research are likely Type I errors, concluding that multiple testing corrections are critical for experimental work~\cite{Ferro2024}.

There are several adjustment procedures for addressing this problem. The common idea behind them is to \emph{adjust} the $p$-values by accounting for the fact that we have several null hypotheses tested. The aim is to control the overall Type I error rate and thus improve the confidence in the decision to reject $H_0$. Several works have advocated the use of adjustment procedures for IR evaluation when comparing more than two systems~\cite{Fuhr2018,Sakai2021,Ferro2024}, but few studies have empirically validated the reliability of these procedures in IR~\cite{Boytsov2013,Ihemelandu2024}. Our work aims to contribute to fill that gap. Boytsov et al.~\cite{Boytsov2013} examined the behaviour of several adjustment procedures using a collection with tens of thousands of topics. They concluded that multiple comparison procedures seem to over-adjust the $p$-values and thus end up with a higher rate of false negatives (i.e. low average power) when having a sample size of 50 queries. Unadjusted procedures seem more attractive since they have fewer false negatives but at the price of having more false positives. More recently, Ihemelandu and Ekstrand~\cite{Ihemelandu2024} employed simulated IR and RecSys data over effectiveness scores and concluded that the best adjustment was Benjamini-Yekutieli since it yielded the lowest Type I error rate and showed great \emph{average} power in mixed-effect-size experiments.

In this work, we use two different approaches to explore the behaviour of multiple comparison adjustments in a typical IR experiment. First, we employ past TREC data to simulate new systems with realistic ranking behaviour by adopting the simulation approach by Parapar et al.~\cite{Parapar2021}. This method learns a model from search system's results and then uses this model to create different experimental situations. We construct scenarios where every system is equal to all others and scenarios where every system is different from the rest. With this certainty about the truth or falseness of the null hypothesis for every pairwise comparison, we can accurately estimate the Type I error rates and \emph{complete} power of the tests. Second, we use real TREC data from the Million Query to further validate our claims. We take the effectiveness scores of a series of systems over hundreds of topics as the ground truth of their differences. By having a large sample size, we may assume that the observed performance is a good estimation of the whole population. Thus, with the effectiveness values from those topics we may decide whether two systems have equal average performance. Then, we sample different subsets of those topics, evaluate the systems only on those samples, and observe which test decisions agree more with the ground truth.

\sloppy We evaluated some of the commonest tests in IR evaluation, namely, the t-test, the Wilcoxon Signed-Rank test, the adjusted two-way ANOVA, and the permutation test for multiple comparisons (randomised TukeyHSD). We used the following adjustment procedures: Bonferroni~\cite{Dunn1961}, Holm~\cite{Holm1979}, Benjamini-Hochberg~\cite{Benjamini1995} and Benjamini-Yekutieli~\cite{Benjamini2001}. Results show that unadjusted tests yield an exceedingly high Type I error rate, emphasising the need to use adjustment procedures when multiple testing. Interestingly, most adjustment procedures do not reach the expected designed ratio of Type I errors. In fact, they produce an error rate lower than expected, especially when adjusting $p$-values yielded by a t-test, although they improve as the number of topics increases. Results on real TREC data show that Wilcoxon plus the Benjamini-Hochberg adjustment is the best in terms of power, and thus, we recommend it for future work.


\section{Background}
\label{sec:rw}

IR experiments commonly involve comparing the performance of several systems by computing effectiveness scores over a series of topics. Statistical tests are required to determine if differences between those scores can be ascribed to something different from random chance. The typical approach in IR is using a two-sided paired test, with the null hypothesis $H_0$  being the equality of mean scores. Given a pair of systems, we compute a test statistic using their per-topic scores and then we compute a $p$-value. This $p$-value quantifies the probability of obtaining a comparable or more extreme test statistic value under the null hypothesis. Based on the $p$-value, we decide whether to reject the null hypothesis of the systems being equal. Rejecting $H_0$ when the $p$-value is below a given threshold $\alpha$ provides us with a way of controlling the Type I error rate. The problem when comparing more than just one pair of systems---which is usually the case---is that we are testing several hypotheses at the same time, which may inflate the Type I error rate. This is known as the multiple comparison problem (MCP). If we have $m$ retrieval systems, comparing all pairs of systems requires $k = \frac{m(m - 1)}{2}$ \emph{different} null hypotheses. The significance level $\alpha$ controls the probability of incorrectly rejecting the null hypothesis when it is true. Thus, the probability of correctly accepting every null hypothesis when they are all true is $(1 - \alpha)^k$. The probability of committing \emph{at least one} Type I error is $1 - (1 - \alpha)^k$. This is the family-wise error rate (FWER)\footnote{Assuming that the $k$ tests are independent.}. For example, if we have $m = 6$ systems, then we would do $k = \frac{6(6 - 1)}{2} = 15$ tests and, with $\alpha = 0.05$, the FWER would be equal to $0.53$. In other words, we would have a 53\% chance of committing one Type I error, i.e. incorrectly identify a result as significant.

\subsection{Controlling the Family-wise Error Rate}

Several strategies have been proposed to control the FWER. Two of the most popular are Bonferroni's and Holm's methods. The common idea is to raise the bar of evidence required to reject any null hypothesis. In other words, these strategies make the reject decision more conservative, which improves confidence in the decision to reject $H_0$.

\myparagraph{Bonferroni Correction.}

The Bonferroni correction adjusts the $p$-value at which every null is rejected based on the total number of tests being performed~\cite{Dunn1961}. Specifically, if we are performing $k$ tests, we only reject those with a $p$-value $\leq \frac{\alpha}{k}$. This very strict correction ensures the FWER is lower than $\alpha$, but being so strict may result in a considerable loss of power.

\myparagraph{Holm's Method.}

Holm's method is also known as Holm's step-down procedure or Holm-Bonferroni method~\cite{Holm1979}. In this case, the threshold that we will use to reject each null hypothesis will depend on the $p$-values of the rest of hypotheses. This is different to the Bonferroni correction, where the threshold only depends on the number of hypotheses being tested. Let $p_1 \leq p_2 \leq ... \leq p_{k-1} \leq p_k$ be the $p$-values of each hypothesis sorted in ascending order, and let $H_i$ be the hypothesis associated with $p_i$. The method proceeds as follows:

\begin{enumerate}
  \item If $p_1 > \frac{\alpha}{k}$, accept all the hypotheses. This is, no difference is significant.

  \item If $p_1 \leq \frac{\alpha}{k}$ reject $H_1$ and consider $H_2$.

  \item If $p_2 > \frac{\alpha}{k - 1}$, accept $H_{i \geq 2}$.

  \item If $p_2 \leq \frac{\alpha}{k - 1}$, reject $H_2$ and consider $H_3$.

  \item Repeat the process until finding the first $j$ such that $p_j > \frac{\alpha}{k - j + 1}$
\end{enumerate}

Other examples of corrections are the Simes-Hochberg~\cite{Simes1986,Hochberg1988}, Hommel~\cite{Hommel1988,Hommel1989} and Rom~\cite{Rom1990} methods. We focus on Bonferroni and Holm since they are the most popular.

These corrections are post-hoc adjustments, i.e., we first compute an unadjusted test and then employ an adjustment procedure to correct the $p$-values. There are another alternatives that already take into account the multiple testing scenario when modelling distributions under the null hypothesis. For example, the two-way ANOVA with TukeyHSD correction and its randomised version, the randomised TukeyHSD test~\cite{Carterette2012}.

\myparagraph{Two-way ANOVA with TukeyHSD correction.} The two-way ANOVA tests the omnibus hypothesis that all the systems compared are statistically equal. Rejecting this hypothesis means that there is one system different from the rest, but we do not know nothing about the pairwise comparisons. Thus, we need to follow-up the ANOVA with a TukeyHSD test, which will give us a $p$-value for each pairwise comparison.

\myparagraph{Randomised TukeyHSD.} This test is a generalisation of the permutation test to multiple comparisons. At each step, the test permutates each array of per-topic scores and then computes the difference between the mean scores for every pair of systems, using the perturbed arrays. The $p$-value for each pair is the number of times that the maximum perturbed difference (i.e. the difference between the maximum score and the minimum score after pertubation) is larger than the observed difference for this pair. We have included a pseudo-code in~\Cref{alg:tukey}.

\begin{algorithm}
  \caption{Paired Randomised Tukey HSD}
  \label{alg:tukey}
  \begin{algorithmic}[1]
    \Input
    \Desc{$X$}{$n\times m$ topic-system scores matrix.}
    \Desc{$B$}{number of permutations.}
    \EndInput
    \Output
    \Desc{$P$}{$n\times n$ matrix holding a p-value for each pairwise system comparison.}
    \EndOutput
    \For{$k \gets 1$ to $B$}
    \State initialise $n \times m$ matrix $X'$
    \For{each topic $t$}
    \State $\textnormal{row t of}~X' \gets \textnormal{permutation of values in row t of}~X$
    \EndFor
    \State $d' \gets \max_{i}\bar{X'_i} - \min_{j}\bar{X'_j}$ \Comment{$\bar{X'_i}$ is the mean of column $i$}
    \For{each pair of systems $i, j$}
    \If{$d' > \lvert \bar{X_i} - \bar{X_j} \rvert$}
    \State $P_{i,j} \gets P_{i,j} + \frac{1}{B}$
    \EndIf
    \EndFor
    \EndFor
  \end{algorithmic}
\end{algorithm}

The main counterpart of these procedures that control the FWER is making the comparison more conservative. This may reduce the power of the test---with respect to unadjusted procedures---, increasing the number of Type II errors\footnote{A Type II error happens when the null hypothesis is incorrectly accepted.}. In fact, there will always be a trade-off between controlling the FWER and the power to reject the null hypothesis. Recall from the previous section that the FWER is defined as the probability of making \emph{at least} one false positive error. Controlling this error makes very unlikely to reject any null hypothesis, and as the number of hypotheses grows, this reduces the power.

\subsection{Controlling the False Discovery Rate}

A recent class of approaches for the MCP focuses on controlling for what is known as the \textit{false discovery rate} (FDR). The FDR is the proportion of false positives among the differences \emph{predicted} as significant~\cite{Benjamini1995}. The rationale is to make the test less strict but still be able to control the ratio of Type I errors.

\subsubsection{Benjamini-Hochberg procedure (BH).}

The first strategy of this kind was proposed by Benjamini and Hochberg~\cite{Benjamini1995}. Let $p_j$ be the $j$-th smallest $p$-value out of the $k$ $p$-values, and let $H_j$ be the hypothesis associated with $p_j$. The FDR $\delta_{j}$ for hypothesis $H_j$ is bounded by $\frac{k \times p_j}{j} \leq \delta_{j}$. Now, if we want an FDR of $\delta$ for the whole experiment, then we must reject all hypotheses that satisfy:

\begin{equation*}
  p_j \leq \delta\frac{j}{k}
\end{equation*}

\subsubsection{Benjamini-Yekutieli procedure (BY).}

Other alternative is the Benjamini-Yekutieli procedure~\cite{Benjamini2001}. This procedure refines the bound used by the BH procedure, such that we must reject the hypotheses that satisfy:

\begin{equation*}
  p_j \leq \delta\frac{j}{k \sum_{i=1}^{k} \frac{1}{i}}
\end{equation*}

It is important to understand the differences between the notions of $\alpha$ and $\delta$. Suppose we control the FDR ($\delta$) for $k$ hypotheses at a level of 0.05. This means that if we repeat the experiment many times, we should expect, \emph{on average}, that 5\% of the rejected null hypotheses will be false positives. Now suppose that we control the FWER ($\alpha$) for $k$ hypotheses at a level of 0.05. This means that we should expect to make at least one false positive 5\% of the times. Here we can see why controlling for the FDR is less stringent than controlling the FWER. In fact, when having just $k= 10$ comparisons (something very likely in IR, since we only need 5 different systems to perform 10 pairwise comparisons), simulations show that controlling the FWER at 0.05 causes the power to fall below 60\%~\cite[Chapter~13]{James2013}.

In this work, we evaluate both adjustments that control the FWER and those that control the FDR.


\section{Method}
\label{sec:method}

We study the behaviour of several multiple comparison procedures by adopting the simulation approach by Parapar et al.~\cite{Parapar2021}. This method allows us to simulate systems with realistic behaviour while controlling the truth or falseness of the null hypothesis. With full control over the null, we can compute accurate error and power rates of the adjustment procedures.

The simulation first fits a regressor for each topic-system pair that exists in a given collection. It uses the relevance data of each document in the given ranking. Using relevance data is less risky and requires fewer assumptions than using score distributions to model the behaviour of a retrieval system~\cite{Parapar2021,Urbano2021,Robertson2013}. This regressor models the appearance of relevant and non-relevant documents on the ranking according to their position. The fitted regressor has this form:

\begin{equation*}
  h_\theta = \frac{1}{1 + e^{-\theta_{0}-\theta_{1} \cdot p}}
\end{equation*}

\noindent where $p$ is the position in the ranking, $\theta_{0}$ and $\theta_{1}$ are the fitted parameters, and $h_\theta$ is the probability of relevance at position $p$.

By sampling from the set of topic regressors, we can generate multiple simulated rankings for the same system to compare the adjustment procedures under $H_0$. We show how this sampling is performed in \Cref{alg:sampling}. Then, by modifying the parameters $\theta_{0}$ and $\theta_{1}$ of each fitted regressor, we can simulate better and worse systems and compare the adjustment procedures under $H_1$. This stochastic simulation models the behaviour of ranking systems in a realistic way, where some queries are improved, and some are degraded, as shown by Parapar et al.~\cite{Parapar2021}. Instead of using this simulation to generate just pairs of two systems~\cite{Parapar2021}, we use this approach to simulate scenarios with multiple comparisons. In particular, we simulate two different scenarios:

\begin{algorithm}
  \caption{Algorithm for sampling.}
  \label{alg:sampling}
  \begin{algorithmic}[1]
    \Input
    \Desc{$h_\theta$}{A fitted logistic regressor.}
    \EndInput
    \Output
    \Desc{$R$}{A simulated ranking.}
    \EndOutput
    \State $R \gets \{\}$;
    \For{$position \gets 1$ to $rank\_size$}
    \State $BernoulliParam \gets h_\theta(position)$
    \State Draw a sample $rel\_position \sim Bernoulli(BernoulliParam)$
    \State $R[position] \gets rel\_position$ \Comment{$rel\_position$ is either 0 or 1}
    \EndFor
  \end{algorithmic}
\end{algorithm}

\subsubsection{Scenario 1: Every Null Hypothesis is True.}

We fit a different regressor for each system-topic pair. We simulate $m$ new rankings for the same system pair by sampling the regressors without modifying their parameters. Since we did not modify the parameters of the regressor, we know we are under $H_0$, and each system is equal to the rest. The steps to simulate systems in this scenario are:

\begin{enumerate}
  \item Given a run with results for every topic, fit a different regressor for each topic-system using the relevance values in the ranking.
  \item Sample $n$ random system regressors $m$ times, thus creating $m$ different system runs from the same regressor, which include rankings for $n$ different topics, without modifying the parameters $\theta_{0}$ and $\theta_{1}$. In this way, the $k = \frac{m (m - 1)}{2}$ null hypotheses are all true.
\end{enumerate}

For each system run, we perform 1000 repetitions of this process to compute accurate Type I error rates. This error rate is the percentage of the 1000 times that the test rejected \emph{at least} one of the $k$ nulls ($p$-value $\leq \alpha$).

\subsubsection{Scenario 2: Every Null Hypothesis is False.}

With the same fitted regressors, we simulate $m$ new system runs while modifying the parameters of the regressors before sampling, so that each run is different from the rest. The steps to simulate system outputs in this scenario are:

\begin{enumerate}
  \item For every run, use the same fitted regressors as before.
  \item Modify the parameters of each regressor by a given proportion $prop$. In particular, if $\theta_i$ is positive we set $\theta^{new}_i = \theta_i \cdot (1 + prop)$. If $\theta_i$ is negative, we set $\theta^{new}_i = \theta_i \cdot \frac{1}{1 + prop}$
  \item Sample each of the $n$ topic-system altered regressor once, thus creating one different run from the same regressor, but with different parameters. This run should be a $prop$\% better in terms of performance. Thus, $H_0$ is false.
  \item Repeat Steps 2 and 3 $m - 1$ times, changing the $prop$ value. Since we are modifying the parameters in every step, the $k$ different null hypotheses are all false.
\end{enumerate}

In this case, we know that the $k = \frac{m (m - 1)}{2}$ null hypotheses are all false because, in every step, we have modified the regressor's parameters.

For each run, we perform 1000 iterations of this process to accurately estimate the power of the tests. In this case, the power is the percentage of the 1000 times a given test rejects all the $k$ hypotheses.


\section{Experiments}
\label{sec:experiments}

We used the simulation methods to evaluate the t-test, the Wilcoxon Signed-Rank test, the Bonferroni and Holm methods that control the FWER, the Benjamini-Hochberg (BH) and Benjamini-Yekutieli (BY) procedures that control the FDR, the two-way ANOVA with TukeyHSD correction, and its randomised version, the randomised TukeyHSD test. We focused on the two-sided paired case, which is common in IR experimentation. We used the data of the TREC-8 and TREC-7 datasets. Results had same trends across both datasets, so we are only reporting those on TREC-8 due to space constraints. This dataset includes 129 runs, where each run is the result of one system for the 50 topics of the collection. We performed 1000 simulations for each topic-system pair, using $m \in \{3, 5, 10\}$ systems and $n \in \{10, 30 , 50\}$ topics. We used Average Precision (AP) as the measure to score the runs. We also experimented with Normalized Discounted Cumulative Gain (NDCG), and the results followed the same trends, so we only reported AP due to space constraints. We set $\alpha = 0.05$ as is common in IR. For computing the randomised TukeyHSD test, we set $B = \num{100000}$ permutations (see \Cref{alg:tukey}). To allow other researchers to reproduce our results, we have released our code\footnote{\url{https://github.com/davidoterof/ecir2025}}.

\myparagraph{Scenario 1: Every Null Hypothesis is True.}

\begin{figure}[t]
  \centering
  \includegraphics[width=0.9\textwidth]{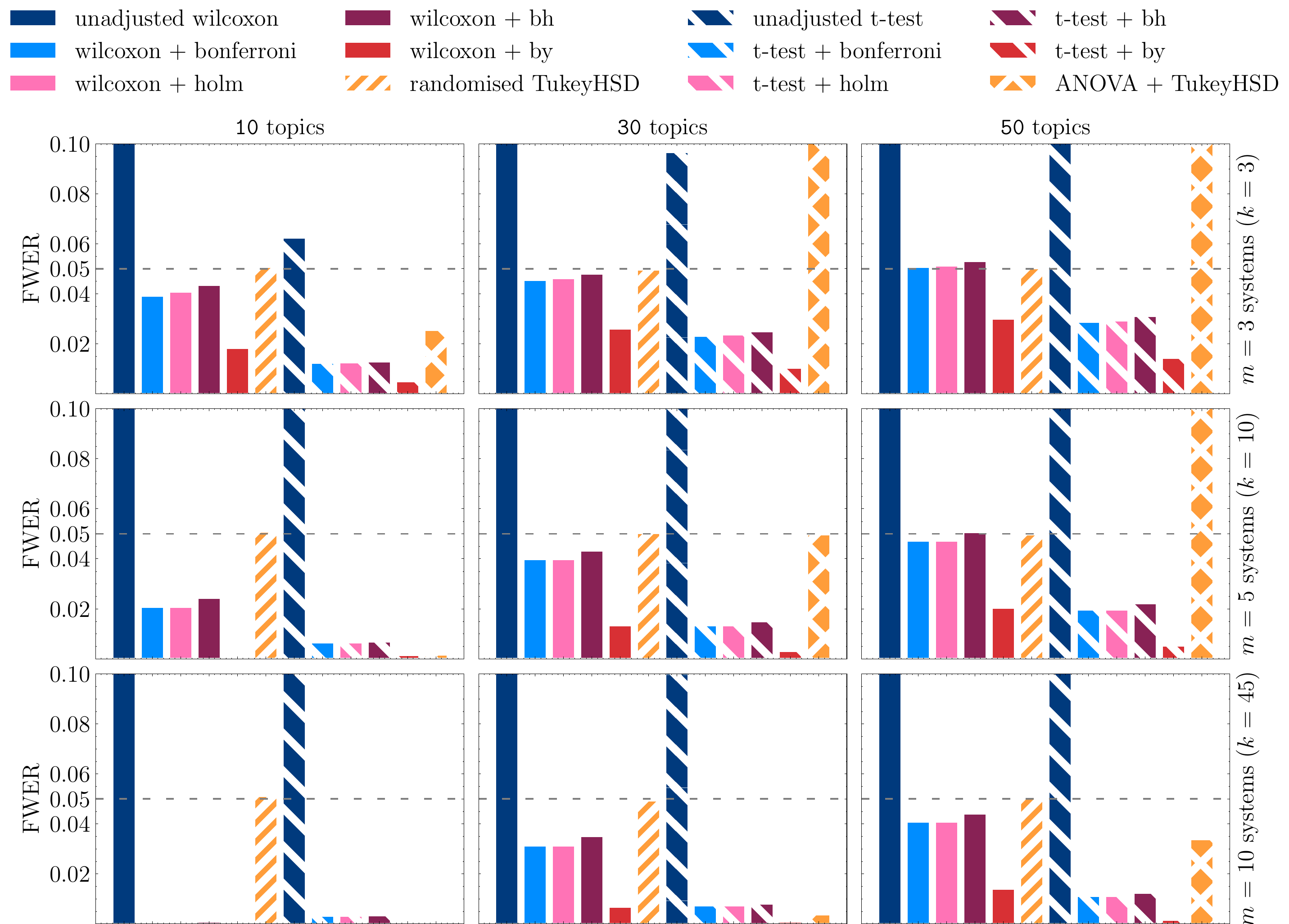}
  \caption{Family-wise error rate of evaluated procedures. Each row is a different number of systems ($k$ is the number of pairwise comparisons). Each column is a different number of topics. The dashed line is the confidence level $\alpha = \delta = 0.05$.}
  \label{fig:error-rate}
\end{figure}

We have summarised the results for the first scenario in \Cref{fig:error-rate}. In this scenario, for each system, we created different runs that are all statistically equal. In other words, for each simulation step, we have a family of null hypotheses that are all true. Thus, we compute the FWER as the number of times that an adjustment procedure marked \emph{at least} one null hypothesis as false. We are plotting this FWER in this figure. Each bar corresponds to the average of $129 \cdot 1000 = \num{129000}$ simulations (1000 simulations per each of the 129 runs of the collection). Plots on the same column correspond to the same topic set size, and plots on the same row correspond to the same number of systems.

First, we observe that these results agree with the theory of the multiple comparison problem: unadjusted procedures have an exceedingly high Type I error rate. In terms of publication bias, performing statistical tests without adjustment may lead researchers to incorrectly identify an advancement when there may be none. Thus, these results serve to further insist on the idea that researchers should employ some adjustment when performing multiple comparisons. In this regard, the \texttt{unadjusted wilcoxon} always yielded a higher error rate than the \texttt{unadjusted t-test}, a finding consistent with previous research~\cite{Parapar2021}.

\sloppy Regarding the adjusted tests, we observe that the \texttt{randomised TukeyHSD} is the only one that matches the expected error rate. The others lag far behind, especially those based on the \texttt{unadjusted t-test}. These other adjustments seem to better match the expected error rate as the number of topics increases. In this regard, in addition to the \texttt{randomised TukeyHSD}, the \texttt{wilcoxon+bh} also matches the expected error rate with 50 topics almost exactly. Interestingly, the BY procedure, which controls the FDR and should be less conservative than Bonferroni and Holm, yields systematically lower Type I error rates. We see some concerning results regarding the ANOVA+TukeyHSD procedure, which yielded very high error rates for 50 topics when the number of systems is small. Overall, these results suggest that the adjustment procedures are over-adjusting the $p$-values, making the tests more conservative than what would be desirable. A test that yields very low error rates may seem more appealing since it means that it is committing fewer false positives. However, as Parapar et al. argued~\cite{Parapar2021}, this means that the adjusted $p$-values do not accurately estimate the probability of finding the observed difference between systems when $H_0$ is true. The confidence level $\alpha$ is a value chosen by the researcher before performing the test, and this value represents the confidence the researcher can have when rejecting $H_0$. A test that does not conform to the expected error rate means that is not accomplishing the design objective of the test. In practice, this has the consequence of the test sacrificing the designed power. Simply put, this means that we are working with a lower \emph{real} confidence level, which shortens the critical region, making less likely to reject the null hypothesis~\cite{Cohen1988}.

\myparagraph{Scenario 2: Every Null Hypothesis is False.}

\begin{figure}[t]
  \centering
  \includegraphics[width=0.9\textwidth]{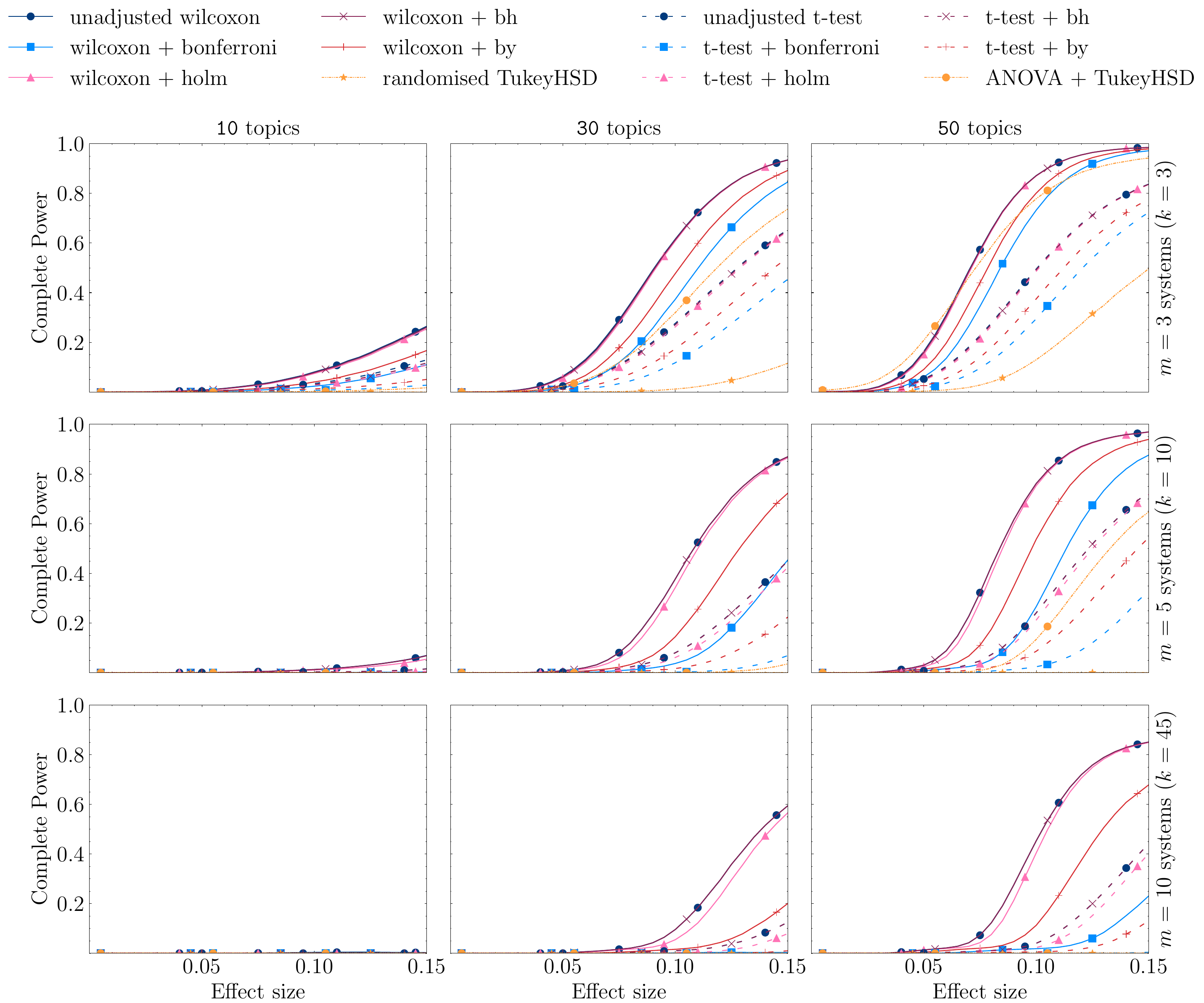}
  \caption{\emph{Complete power} of evaluated procedures. Each row is a different number of systems ($k$ is the number of pairwise comparisons). Each column is a different number of topics. Solid lines correspond to wilcoxon-based tests. Dashed lines correspond to t-test-based tests. This figure is better viewed in colour.}
  \label{fig:power}
\end{figure}

We present results for the second scenario in \Cref{fig:power}. In this scenario, every null hypothesis is false, that is, every comparison between any pair of systems should be flagged as significant. We measure the \emph{complete} power, i.e., the probability to reject every null hypothesis when they are all false, as in this case. There are alternative definitions of power: minimal power, the probability of rejecting at least one null hypothesis; and average power, the ratio of rejected null hypotheses. However, we believe that complete power is of more interest for IR evaluation, since it is common to find several pairs actually different systems when multiple testing in IR.

We observe that with few topics tests have almost no power. This goes in line with the already known result that more topics provide more power~\cite{Carterette2008,Sanderson2005,Webber2008}. The \texttt{wilcoxon + bh} and the \texttt{unadjusted wilcoxon} are clear winners, since they yield higher power values in every case. We also observe that, in general, the most powerful adjustment procedures are those that better met the Type I expected error rate in the previous experiment. Thus, this allows us to conclude that being unable to meet the proportion of expected Type I errors results in a loss of power. Additionally, we see that adjustment procedures do not generally result in a loss of power with respect to unadjusted tests. Interestingly, the \texttt{randomised TukeyHSD}, a test recommended and used in other works~\cite{Carterette2012,Sakai2018}, and that almost perfectly matched the expected Type I error rate, has very low power. This test is very conservative in considering a result as significant because the observed difference has to be larger than the difference between the maximum perturbed mean and the minimum perturbed mean (see \Cref{alg:tukey}). Since we have to perform many permutations ($B$) for the algorithm to converge, many of these permutations will likely yield a difference larger than the observed one, thus increasing the $p$-values. This effect becomes even more severe as the number of hypotheses tested grows. As we see in \Cref{fig:power}, with 10 and 45 hypotheses (second and third rows), this test has virtually no power. Thus, a researcher employing this test will likely fail to detect results that are indeed significant.

\myparagraph{Experiments on the Million Query Data.}

As we have seen, our simulation framework allowed us to study the behaviour of several adjustment procedures for multiple testing under common IR evaluation setups. The validity of this simulation to generate realistic ranking behaviour was already demonstrated in the past~\cite{Parapar2021}. They showed how simulated rankings had strong ranking correlations with real systems and also tested the similarity between the two AP distributions (real TREC system vs. simulated one) and found no noticeable difference. However, we acknowledge that relying only on simulated data might be risky. Some researchers pointed out that this way of simulating per-query improvements, which allowed us a comparison where $H_0$ was false, although stochastic, might not be realistic with respect to how real retrieval systems behave. For this reason, we also experimented with the data of the Million Query Track of 2009~\cite{Carterette2009b}, following the same approach as Boytsov et al.~\cite{Boytsov2013}, to further evaluate our claims under $H_1$. This collection includes 35 runs and 687 topics with judgements. We use the average AP scores of the runs on the 687 topics as a proxy of long-term performance. We set a threshold $\gamma = 0.05\%$, and consider every pair of systems with a MAP difference---over 687 topics---larger than $\gamma$ as different. In this way, we establish the relative performance between every pair of systems. This resulted in 590 pairs of systems that were actually different---according to our criterion---among the 595 possible pairs. For the 5 lefout pairs, we cannot conclude they are actually equal. The fact that they do not have a difference larger than the 0.05\% does not mean that they are equal. In other words, we do not have enough evidence to state that they are actually equivalent. We centre our analysis on the comparisons where $H_0$ is false---the pairs that are different--- and we evaluate the average power. In particular, we want to evaluate if the output of the tests with fewer topics agrees with the ground truth. To this aim, given a sample size, we sampled different subsets of topics and measured the performance of the runs on each subset. We carried out 2000 iterations of the sampling process and performed unadjusted and adjusted tests to assess if systems' differences on these subsets were significant. We evaluated if the output of the tests in the sampled subsets agrees with the ground truth.

\begin{figure}[t]
  \centering%
  \includegraphics[width=0.9\textwidth]{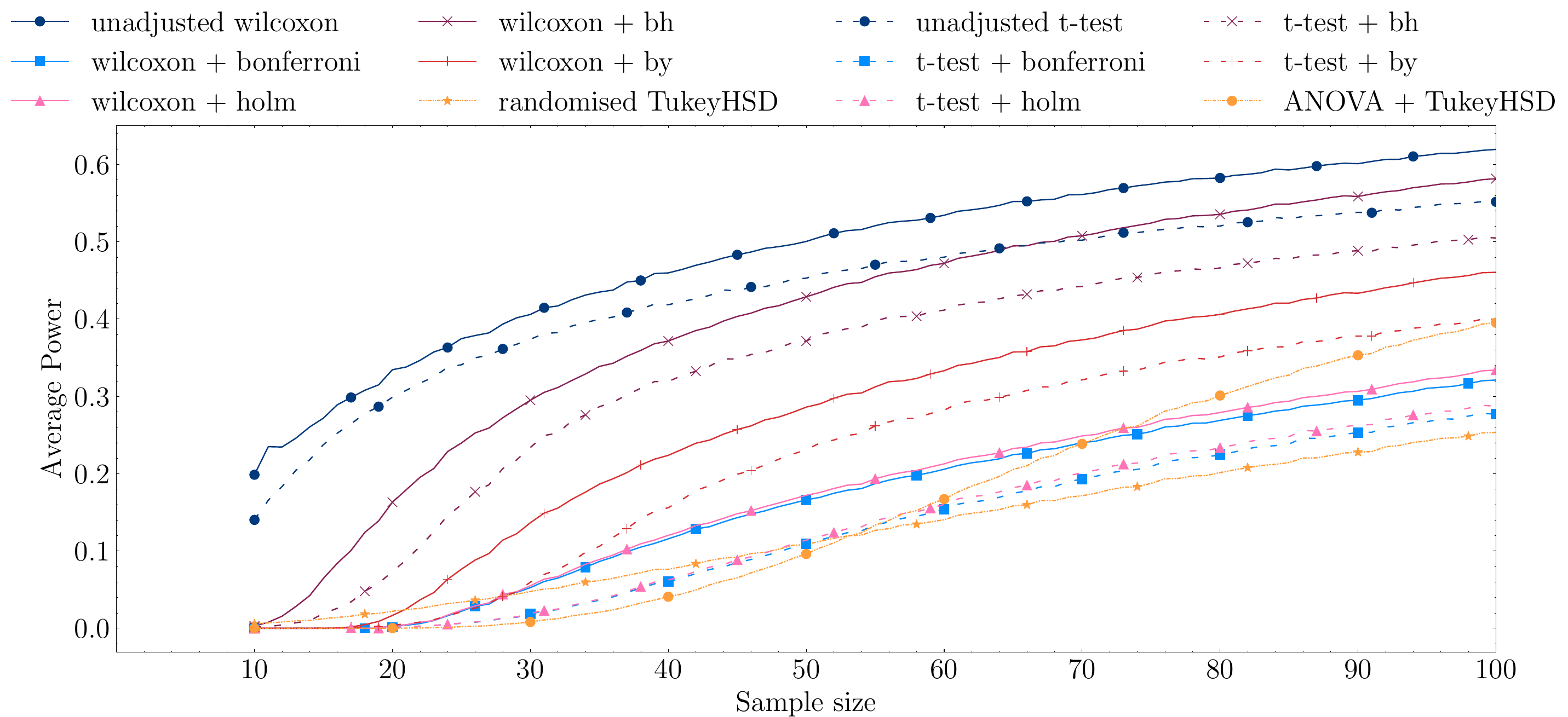}%
  \caption{\emph{Average} power observed in the Million Query experiment. For each sample size we performed 2000 sampling iterations. Solid lines correspond to wilcoxon-based tests. Dashed lines correspond to t-test-based tests. Image is better viewed in colour.}%
  \label{fig:mq}%
\end{figure}

We show the results of this experiment in \Cref{fig:mq}. The average power is the ratio of correctly rejected null hypotheses. From these results, we highlight several findings. Regarding the unadjusted tests, \texttt{wilcoxon} has more power than the \texttt{t-test} for every sample size, something we already observed in our simulation experiment. This result is consistent with past research~\cite{Parapar2021}. Regarding the adjusted tests, the BH procedure is the most powerful, particularly when applied over an unadjusted Wilcoxon, agreeing again with results on simulated data. We also highlight that the \texttt{randomised TukeyHSD}, which yields remarkably good results in terms of expected FWER, performs poorly in terms of power. For sample sizes larger than 50, is indeed the worst test. In light of these results, we recommend researchers to avoid employing this test, since it likely leave out interesting and significant results.


\section{Conclusions}
\label{sec:conclusions}

Several works have advocated for using adjusted tests for IR evaluation when performing multiple tests to avoid publishing supposedly significant results when they were just due to chance. We have empirically investigated the behaviour of several alternatives for controlling the error rates in this kind of scenario. To this aim, we used realistic TREC simulated data that allowed us to have complete control over the truth or falseness of the null hypothesis. We computed accurate error and power rates under different scenarios with different sample sizes. Our results show that, indeed, multiple testing without any adjustment drastically increases the Type I error rate. This result serves to further insist on the idea of emplolying some adjusted test when multiple testing in IR evaluation. In terms of Type I errors, only one procedure, the randomised TukeyHSD, behaved as expected, controlling the expected error rate for every topic set size we tested. The rest of the procedures, especially when adjusting p-values yielded by a two-sided t-test, showed error rates much lower than the expected. Although this may seem better from a practical perspective, we have argued that a test that does not agree with the expected error rate may not be appropriate and may result in a loss of power. This effect is alleviated as the number of topics increases. In particular, the BH procedure is the one that benefits most from the increase in the number of topics, matching the expected error rate almos exactly with 50 topics. We also showed, both with simulated and real data from the Million Query Track, that adjusted tests do not cause a loss of power compared to the unadjusted ones since we observed that they yield virtually the same power for any sample size and any number of hypotheses we evaluated. Among the adjusted tests, the \texttt{wilcoxon+bh} yielded the highest power in every case.

Overall, our approach allowed us to pinpoint which test works better in different experimental conditions, i.e. different number of topics and different number of systems under comparison. Regarding the question of which test to use for IR experimentation, we recommend a Wilcoxon Sign Rank test and employing the Benjamini-Hochberg procedure to adjust the $p$-values. In our experiments, this combination was always the second best in terms of Type I errors. In terms of power, it is the most powerful combination after the unadjusted tests.

\begin{credits}
\subsubsection{\ackname} The authors thank the financial support supplied by the Consellería de Cultura, Educación, Formación Profesional e Universidades (accreditation 2019-2022 ED431G/01, ED431B 2022/33) and the European Regional Development Fund, which acknowledges the CITIC Research Center in ICT as a Re- search Center of the Galician University System and the project PID2022-137061OB-C21 (Ministerio de Ciencia e Innovación supported by the European Regional Development Fund). The authors also thank the funding of project PLEC2021-007662 (MCIN/AEI/10.13039/501100011033, Ministerio de Ciencia e Innovación, Agencia Estatal de Investigación, Plan de Recuperación, Transformación y Resiliencia, UniónEuropea-Next Generation EU).

\end{credits}

\bibliography{references}

\begin{thebibliography}{10}
\providecommand{\url}[1]{\texttt{#1}}
\providecommand{\urlprefix}{URL }
\providecommand{\doi}[1]{https://doi.org/#1}

\bibitem{Benjamini1995}
Benjamini, Y., Hochberg, Y.: Controlling the false discovery rate: A practical
  and powerful approach to multiple testing. Journal of the Royal Statistical
  Society: Series B (Methodological)  \textbf{57}(1),  289--300 (1995).
  \doi{10.1111/j.2517-6161.1995.tb02031.x}

\bibitem{Benjamini2001}
Benjamini, Y., Yekutieli, D.: The control of the false discovery rate in
  multiple testing under dependency. The Annals of Statistics  \textbf{29}(4),
  1165--1188 (2001). \doi{10.1214/aos/1013699998}

\bibitem{Boytsov2013}
Boytsov, L., Belova, A., Westfall, P.: Deciding on an adjustment for
  multiplicity in ir experiments. In: Proceedings of the 36th International ACM
  SIGIR Conference on Research and Development in Information Retrieval. pp.
  403--412. Association for Computing Machinery, New York, NY, USA (2013).
  \doi{10.1145/2484028.2484034}

\bibitem{Carterette2012}
Carterette, B.: Multiple testing in statistical analysis of systems-based
  information retrieval experiments. ACM Transactions on Information Systems
  \textbf{30}(1) (2012). \doi{10.1145/2094072.2094076}

\bibitem{Carterette2008}
Carterette, B., Pavlu, V., Kanoulas, E., Aslam, J.A., Allan, J.: Evaluation
  over thousands of queries. In: Proceedings of the 31st Annual International
  ACM SIGIR Conference on Research and Development in Information Retrieval.
  pp. 651--658. Association for Computing Machinery, New York, NY, USA (2008).
  \doi{10.1145/1390334.1390445}

\bibitem{Carterette2009b}
Carterette, B., Pavlu, V., Fang, H., Kanoulas, E.: Million query track 2009
  overview. In: Proceedings of The Eighteenth Text REtrieval Conference, {TREC}
  2009, Gaithersburg, Maryland, USA, November 17-20, 2009. {NIST} Special
  Publication, vol. 500--278. National Institute of Standards and Technology
  {(NIST)} (2009),
  \url{http://trec.nist.gov/pubs/trec18/papers/MQ09OVERVIEW.pdf}

\bibitem{Cohen1988}
Cohen, J.: Statistical Power Analysis for the Behavioral Sciences. Lawrence
  Erlbaum Associates (1988)

\bibitem{Dunn1961}
Dunn, O.J.: Multiple comparisons among means. Journal of the American
  Statistical Association  \textbf{56}(293),  52--64 (1961).
  \doi{10.1080/01621459.1961.10482090}

\bibitem{Ferro2022}
Ferro, N., Sanderson, M.: How do you test a test? a multifaceted examination of
  significance tests. In: Proceedings of the Fifteenth ACM International
  Conference on Web Search and Data Mining. pp. 280--288. WSDM '22, Association
  for Computing Machinery, New York, NY, USA (2022).
  \doi{10.1145/3488560.3498406}

\bibitem{Ferro2024}
Ferro, N., Sanderson, M.: Uncontextualized significance considered dangerous.
  In: Proceedings of the 47th International ACM SIGIR Conference on Research
  and Development in Information Retrieval. pp. 261--270. Association for
  Computing Machinery, New York, NY, USA (2024). \doi{10.1145/3626772.3657827}

\bibitem{Fuhr2018}
Fuhr, N.: Some common mistakes in {IR} evaluation, and how they can be avoided.
  SIGIR Forum  \textbf{51}(3),  32--41 (2018). \doi{10.1145/3190580.3190586}

\bibitem{Hochberg1988}
Hochberg, Y.: A sharper bonferroni procedure for multiple tests of
  significance. Biometrika  \textbf{75}(4),  800--802 (1988).
  \doi{10.2307/2336325}

\bibitem{Holm1979}
Holm, S.: A simple sequentially rejective multiple test procedure. Scandinavian
  Journal of Statistics  \textbf{6}(2),  65--70 (1979),
  \url{https://www.jstor.org/stable/4615733}

\bibitem{Hommel1988}
Hommel, G.: A stagewise rejective multiple test procedure based on a modified
  bonferroni test. Biometrika  \textbf{75}(2),  383--386 (1988)

\bibitem{Hommel1989}
Hommel, G.: A comparison of two modified bonferroni procedures. Biometrika
  \textbf{76}(3),  624--625 (1989). \doi{10.1093/biomet/76.3.624}

\bibitem{Ihemelandu2024}
Ihemelandu, N., Ekstrand, M.D.: Multiple testing for ir and recommendation
  system experiments. In: Proceedings of 46th European Conference on IR
  Research (ECIR 2024). pp. 449--457. ECIR '24, Springer Nature Switzerland
  (2024)

\bibitem{James2013}
James, G., Witten, D., Hastie, T., Tibshirani, R.: An Introduction to
  Statistical Learning: with Applications in R (2013)

\bibitem{Parapar2021}
Parapar, J., Losada, D.E., Barreiro, A.: Testing the tests: Simulation of
  rankings to compare statistical significance tests in information retrieval
  evaluation. In: Proceedings of the 36th Annual ACM Symposium on Applied
  Computing. pp. 655--664. SAC '21, Association for Computing Machinery, New
  York, NY, USA (2021). \doi{10.1145/3412841.3441945}

\bibitem{Parapar2020}
Parapar, J., Losada, D.E., Presedo-Quindimil, M.A., Barreiro, A.: Using score
  distributions to compare statistical significance tests for information
  retrieval evaluation. Journal of the Association for Information Science and
  Technology  \textbf{71}(1),  98--113 (2020). \doi{10.1002/asi.24203}

\bibitem{Robertson2013}
Robertson, S., Kanoulas, E., Yilmaz, E.: Modelling score distributions without
  actual scores. In: Proceedings of the 2013 Conference on the Theory of
  Information Retrieval. pp. 85--92. Association for Computing Machinery, New
  York, NY, USA (2013). \doi{10.1145/2499178.2499181}

\bibitem{Rom1990}
Rom, D.M.: A sequentially rejective test procedure based on a modified
  bonferroni inequality. Biometrika  \textbf{77}(3),  663--665 (1990),
  \url{http://www.jstor.org/stable/2337008}

\bibitem{Sakai2018}
Sakai, T.: Laboratory Experiments in Information Retrieval, The Information
  Retrieval Series, vol.~40. Springer (2018). \doi{10.1007/978-981-13-1199-4}

\bibitem{Sakai2021}
Sakai, T.: On {F}uhr's guideline for {IR} evaluation. SIGIR Forum
  \textbf{54}(1) (2021). \doi{10.1145/3451964.3451976}

\bibitem{Sanderson2005}
Sanderson, M., Zobel, J.: Information retrieval system evaluation: Effort,
  sensitivity, and reliability. In: Proceedings of the 28th Annual
  International ACM SIGIR Conference on Research and Development in Information
  Retrieval. pp. 162--169. SIGIR '05, Association for Computing Machinery, New
  York, NY, USA (2005). \doi{10.1145/1076034.1076064}

\bibitem{Simes1986}
Simes, R.J.: An improved bonferroni procedure for multiple tests of
  significance. Biometrika  \textbf{73}(3),  751--754 (1986).
  \doi{10.2307/2336545}

\bibitem{Smucker2007}
Smucker, M.D., Allan, J., Carterette, B.: A comparison of statistical
  significance tests for information retrieval evaluation. In: Proceedings of
  the Sixteenth ACM Conference on Conference on Information and Knowledge
  Management. pp. 623--632. CIKM '07, Association for Computing Machinery, New
  York, NY, USA (2007). \doi{10.1145/1321440.1321528}

\bibitem{Urbano2021}
Urbano, J., Corsi, M., Hanjalic, A.: How do metric score distributions affect
  the type i error rate of statistical significance tests in information
  retrieval? In: Proceedings of the 2021 ACM SIGIR International Conference on
  Theory of Information Retrieval. pp. 245--250. Association for Computing
  Machinery, New York, NY, USA (2021). \doi{10.1145/3471158.3472242}

\bibitem{Urbano2019}
Urbano, J., Lima, H., Hanjalic, A.: Statistical significance testing in
  information retrieval: An empirical analysis of type {I}, type {II} and type
  {III} errors. In: Proceedings of the 42nd International ACM SIGIR Conference
  on Research and Development in Information Retrieval. pp. 505--514. SIGIR'19,
  Association for Computing Machinery, New York, NY, USA (2019).
  \doi{10.1145/3331184.3331259}

\bibitem{Webber2008}
Webber, W., Moffat, A., Zobel, J.: Statistical power in retrieval
  experimentation. In: Proceedings of the 17th ACM Conference on Information
  and Knowledge Management. p. 57100580. CIKM '08, Association for Computing
  Machinery, New York, NY, USA (2008). \doi{10.1145/1458082.1458158}

\end{thebibliography}
\bibliographystyle{splncs04}

\end{document}